\documentclass{aa}
\usepackage{natbib}
\usepackage{txfonts}
\usepackage{graphicx}
\usepackage{endnotes}
\def\diff{{\mathrm d}}
\def\imag{{\rm i}}  
\def\msol{\mathrm{M}_\odot}

\def\mast{M_\ast}

\def\msol{\mathrm{M}_\odot}
\def\lsol{\mathrm{L}_\odot}

\def\rast{{\sl R}_\ast}

\def\llsol{L/\mathrm{L}_\odot}

\def\epst{\varepsilon_{T}}

\def\cp{c_P}
\def\signorm{\sqrt{3 G M_\ast / R_\ast^3}}

\def\sigi{\sigma_{\rm I}}
\begin{document}

   \title{Oscillatory Secular Modes: The Thermal Micropulses}

   \author{A.Gautschy \inst{1} and L. G. Althaus \inst{2}}

   \institute{Wetterchr\"uzstr. 8c, 4410 Liestal, Switzerland     
            \\
              \and
              Facultad de Ciencias Astron{\'o}micas y Geof{\'i}sicas,
              Universidad Nacional de La Plata, Paseo del Bosque
              S/N (1900) La Plata and 
              Instituto de Astrofisica La Plata (IALP), Argentina}

   \date{Received date .......; accepted date .......}

%
\abstract{}
{Stars in the narrow mass range of about $2.5$ and $3.5 \msol$ can
develop a thermally unstable He-burning shell during its ignition
phase. We study, from the point of view secular stability theory,
these so called thermal micropulses and we investigate their
properties; the thermal pulses constitute a convenient conceptual
laboratory to look thoroughly into the physical properties of a
helium-burning shell during the \emph{whole} thermally pulsing
episode.} {Linear stability analyses were performed on a large number
of $3 \msol$ star models at around the end of their core
helium-burning and the beginning of the double-shell burning
phase. The stellar models were not assumed to be in thermal
equilibrium.}  {The thermal mircopulses, and we conjecture all other
thermal pulse episodes encountered by shell-burning stars, can be
understood as the nonlinear finite-amplitude realization of an
oscillatory secular instability that prevails during the whole thermal
pulsing episode. Hence, the cyclic nature of the thermal pulses
can be traced back to a linear instability concept.}  {}

\titlerunning{Oscillatory Secular Modes}
\maketitle
\section{Introduction}
\label{sect:intro}
According to the published literature,
\citet{guredinski47} seem to have been the first
\footnote{The paper referred to in Zel'dovich \& Novikov's ``Stars
and Relativity'' is a ``translation'' for astronomers that appeared in
$1955$ in {\it Trudy chetvertogo sovshchaniya po voprosam kosmogonii},
$143$. The early work by Gurevich \& Lebedinski seems to have made its
way to the west essentially through the reference in the Zel'dovich \&
Novikov text, as whenever astronomers refer to Gurevich \&
Lebedinksi, they usually cite the $1955$ version and not the technical
original of $1947$.}
to study the \emph{thermal stability of nuclear-burning shells} in
stars; the authors aimed at revealing the physical nature of novae and
supernovae. The work of Gurevich \& Lebedinksi apparently went
unnoticed outside of the Soviet Union so that $15$ years passed before
the field advanced: The seminal paper of
\citet{schwarm65} reported the thermal instability of the He-burning
shell in a $1 \msol$ asymptotic giant-branch (AGB) star.  The
instability was stumbled upon once the time-step of the evolution
computation was chosen sufficiently small during the pertinent
evolutionary phase.  Independently of Schwarzschild \& H\"arm,
\citet{weigert65, weigert66}
\footnote{The $5\msol$ star studied by Weigert was previously
computed into the advanced evolutionary stage by Kippenhahn (see
\citet{kippenhahn65}). He could not get the star to eventually
turn into a white dwarf as a carbon flash developed on the AGB.  The
star's center heated up too much as neutrino cooling was
neglected. Weigert's computations included neutrino energy
losses; this energy sink kept the central temperature low enough to
prevent the onset of central carbon burning and the star evolved
further along the AGB to finally run into the thermal instability of
the He burning shell.}
observed the same phenomenon at the beginning of the double-shell
burning phase of a $5\msol$ star model.  Weigert's computations
indicated already the cyclic nature of the
instability. \citet{rose66}, studying the advanced evolution of a
$0.53 \msol$ helium star, also picked up thermal pulses once the
evolutionary time-step became sufficiently short. As Rose was
interested in the evolution to the white-dwarf domain of He-star
models, he chose large time-steps across the helium shell burning
phase so that he captured only the last three thermal pulse cycles of
the instability episode. In any case, \citet{rose66} emphasized, as
Weigert did, the cyclic nature of the phenomenon and he was first to
present eigenfunctions in the surroundings of the unstable nuclear
burning shell.

Already \citet{Ledoux1962} noted that, in principle at least, the
secular stability problem admits of complex eigensolutions.  The
stability analysis of thin nuclear-burning shells was extended in
\citet{haerschild72} wherein the thermal pulses, eventually considered 
as a cyclic phenomenon, were associated with the star's secular
modes. The authors connected the \emph{onset} of the thermal
pulses with an overstable \emph{oscillatory secular mode}. After the
first few pulse cycles, however,``[\dots] the oscillatory instability
gives way [\dots] to a simple exponential instability which causes the
repetitive helium-shell flashes described in earlier investigations
[\dots]" as \citet{haerschild72} put it. This last statement seems to
have been taken as the final word on the issue and was perpetuated as
such in the astronomical literature thereafter. 

Despite the apparent settlement of the case, the point of view adopted
by \citet{haerschild72} continued to intrigue as it does not explain
the cause of the \emph{cyclic} nature of the thermal pulses once the
oscillatory instability gives way to an exponential one. The numerical
studies showed that the period of the initial oscillatory secular mode
was close to the length of the developing thermal pulse cycles of the
helium shell. Why or how the star remembered the linear result so
closely remained unexplained \citep{hansen78}.  If the cyclic nature
of the thermal pulses is indeed a nonlinear phenomenon that escapes
accessibility of a linear analysis, then the ``relaxation
oscillations'' of the thermal-pulse phase are eventually one case of the
realization of \emph{hard self-excited oscillations}
\citep{jpc80} in stars \citep[see also][]{buchlerperdang79}.

Besides the thermal pulses that are associated with secularly unstable
thin helium-burning shells in low- and intermediate-mass stars along
the asymptotic giant branch, secularly unstable shell burning is also
encountered at the onset of off-center helium burning in low-mass
stars as they evolve from the tip of the red-giant branch onto the
horizontal branch \citep[e.g.][]{hcthomas67, despain81}.  But also
more exotic episodes of stars' lives can get entangled in He-shell
instabilities: X-ray bursts as observed in mass-accreting neutron
stars \citep[e.g.][]{hansenvanhorn75, bildsten95}. Even
hydrogen-burning shells on accreting white dwarfs can go cyclically
unstable \citep[e.g.][and references in the
latter]{giannoneweigert67,cassisietal98} via the same physical
mechanism. Hence, a solid understanding of the thermal pulses is
beneficial for disparate fields of stellar astrophysics.

\medskip

This paper deals with those thermal pulses that are encountered when,
under suitable circumstances, a helium shell builds up and ignites at
the end of core He burning of a $3\msol$ star
\citep{mazzitona86}. Applying the simple, local instability criterion
for thin nuclear-burning shells of \citet{schwarm65},
\citet{mazzitona86} concluded that the instability was physical and
due to the same mechanism as encountered later, higher up along the
AGB. The thermal pulses at the onset of He shell burning were termed
\emph{thermal micropulses} (ThMPs) as the surface luminosity varies
only marginally as compared to the variation during a thermal pulse on
the advanced AGB.  In an attempt to evolve a star from the main
sequence to the terminal Debye-cooling stage as a white dwarf, thermal
micropulses were also encountered by \citet{althausetalthmps02} in
stellar models for which semiconvection was neglected in the
computations. It seems that only stars in the narrow mass range of
$2.5 - 3.5 \msol$ develop ThMPs under suitable conditions.

The evolutionary tracks of stars undergoing ThMPs
\citep[cf.][]{mazzitona86,althausetalthmps02} show that the
manifestation of the pulses in the stars' observables are minimal. The
pulse amplitudes in the total luminosity are small and the time-scale
of the pulses so short that effects on a population of stars on the
early second ascent of the giant branch are very unlikely to be
detectable even in large stellar aggregates. Despite ThMPs being
unlikely to be observable or possibly even realized in nature, their
very existence in \emph{star models} constitutes a welcome
opportunity: With the ThMPs, we can theoretically study the stability
properties of a secularly unstable nuclear-burning shell from onset of
the instability to the very end when the shell stabilizes itself
again. In particular, the self-stabilization of the shell
distinguishes the ThMPs from thermal pulses along the upper
AGB. Furthermore, we expect that the properties of the secular mode
spectrum of the ThMP phase are also representative for the thermally
pulsing stars on the advanced AGB.

In the following, we use the ThMP phase as a idealized laboratory
to study the episode of cyclic thermal pulses once again from the
point of view of linear secular stability theory.  We take advantage
of the result of \citet{gabriel72} that over time-scales that are
short compared with a star's nuclear time-scale its quasi-static
evolution can be expanded in terms of secular eigendata.
  
We follow the pertinent part of the secular eigenspectrum throughout the
\emph{whole} ThMP episode and we show that ThMPs can be understood as
an oscillatory secular instability; hence nonlinear theory via
relaxation oscillations does not have to be invoked to explain
the existence of the cyclic instability. Nonlinear effects, on the
other hand, determine the temporal shape of the observables of the
pulses and they seem to govern the amplitude evolution of the pulse
cycles.

\begin{figure}
      \resizebox{.98\hsize}{!} {\includegraphics{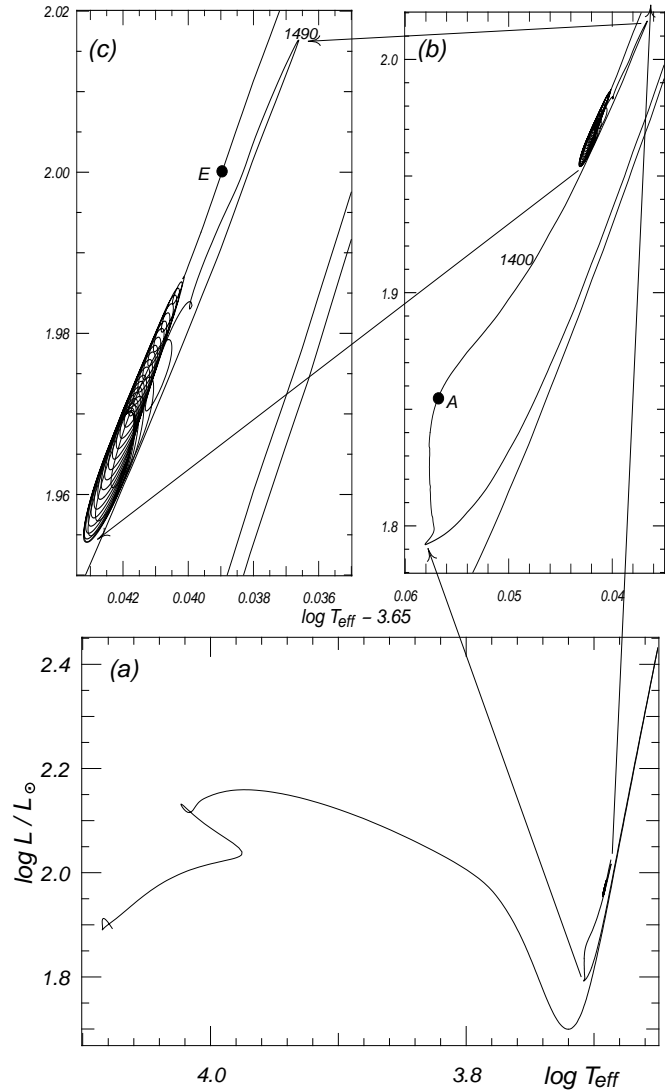}}
      \caption{Evolutionary track of a $3 \msol$ star with pertinent
      zoom-ins to the epochs of core He burning (panel b) and to the
      development of double-shell nuclear burning (panel c).  Core He
      burning starts at the tip of the first giant branch that is seen
      in the upper right of panel a. The transition from He burning in
      the core to He burning in a shell leads to a temporary
      luminosity drop (after model $1490$ which is designated in panel
      c); during this luminosity decrease, the thermal micropulses set
      in. After the He-burning shell is well developed, the
      micropulses die out and the star continues its second ascend of
      the giant branch towards the asymptotic giant branch. The epochs
      of the begin and the end of the linear secular stability
      analyses are indicated by the letters A and E, respectively. The
      location of model numbers $1400$ and $1490$ along the
      evolutionary track are given for later reference in the text.  }
\label{fig:m30tracks}
\end{figure}

\section{Modeling Procedure}
\label{sect:modelling}

Stellar evolution was computed with the LPCODE code described in
\citet{althausetalthmps02} and \citet{althausetal03}. 
Convective overshooting was modeled as an exponentially decaying
diffusive process above and below every convective shell, including
the boundary of the convective core (during the main-sequence and
central helium-burning phase) and the convective envelope \citep[for
details see][]{althausetal03}.  The nuclear network considered for the
stellar modeling accounts explicitly for $16$ chemical elements and
$34$ thermonuclear reaction rates to follow hydrogen and helium
burning.  Abundance changes of the nuclear species were included
by means of a time-dependent scheme that simultaneously treats nuclear
evolution and mixing processes.  The efficiency of convective mixing
was described by appropriate diffusion coefficients which are
specified by the treatment of convection. The equation of state
included partial ionization, ionic contributions, partially degenerate
electrons, and Coulomb interactions.  Radiative opacities, including
carbon- and oxygen-rich compositions, were those from the OPAL sets
\citep{opal96}, complemented, at low temperatures, with the molecular
opacities from \citet{alexanderferguson94}.

The computations to investigate the linear secular stability
properties of the star models were performed with the Riccati code for
radial nonadiabatic stellar pulsations; the current version of the
code is based on the one described in \citet{gagla90a}.  In addition
to the canonical radial pulsation equations, we used also modified
stability equations wherein the condition of thermal equilibrium of
the background model was relaxed.  For completeness, the relevant
linear stability equations are presented in Appendix A.

\section{Results}
\label{sect:results}
This section contains the results obtained from the computation of the
evolution of a $3 \msol$ model with initially homogeneously
distributed abundances of $X = 0.705$, $Y=0.275$. We followed the
star's evolution from the main sequence to stable double-shell
burning on the lower AGB.  The second part of this section is devoted
to the secular stability analyses performed on the evolutionary star
models between about the ignition of the helium-burning shell and the
termination of the thermal pulses.

\subsection{The micropulses as seen in stellar evolution computations}
\label{sect:thmps_evol}
\begin{figure}
      \resizebox{.98\hsize}{!} {\includegraphics{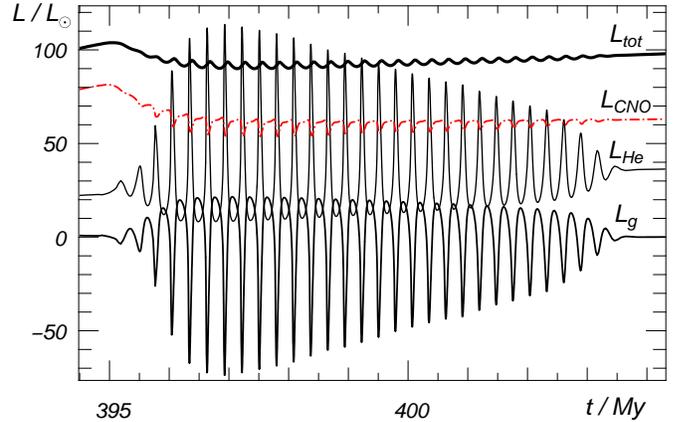}}
      \caption{Variation of various luminosities during the
      thermal micropulse phase of the evolution of a $3 \msol$
      model. $L_{\rm{tot}}$ stands for the total stellar luminosity;
      it is composed of the helium-burning one ($L_{\rm{He}}$), the
      luminosity generated in the hydrogen shell ($L_{\rm{CNO}}$), and
      the luminosity induced by expansion/contraction ($L_{\rm{g}}$)
      in the own gravitational field. The abscissa measures the
      age of the star in $10^6$~years (My).}
\label{fig:m30dpulses}
\end{figure}

Figure~\ref{fig:m30tracks} shows, on three zoom levels, the evolution
on the Hertzsprung-Russell (HR) plane of the $3 \msol$ star mentioned
before.  The displayed track resulted from the computations without
convective overshooting~--~thermal micropulses developed only in such 
stellar models; this aspect will be returned to further
down. Panel (a) captures the star's locus between core hydrogen
burning on the main sequence and the establishing of two burning
shells at the bottom of the AGB.  Core helium burning sets in at the
tip of the first giant branch at the upper right of panel (a). Panel
(b) shows the stump of a blue loop that develops during core helium
burning. After passing through a local luminosity minimum at $\log
\llsol = 1.792$, with a central helium abundance $Y_{\mathrm{c}} =
0.8503$, the star's luminosity rises again to pass through a second
local luminosity maximum ($\log\llsol = 2.015$). Label A denotes the
epoch of the evolution at which the secular stability analyses were
started. For ease of presentation, the secular stability results will
be displayed mostly as a function of model number; label A corresponds
to model number $1300$ with $Y_{\mathrm{c}} = 0.288$. The helium shell
ignited at $Y_{\mathrm{c}} = 0.04$; the corresponding model number,
$1400$, is labeled along the track in panel (b).  At the second local
luminosity maximum (model number $1490$ in panel c),
$Y_{\mathrm{c}}$ dropped to $0.0325$ and the triple-$\alpha$
energy-generation rate of the nuclear-burning shell reached the same
strength as that of core helium burning. Hence, after model 1490,
the star's helium burning is shell dominated. Finally, panel (c) zooms
in to the cyclically varying luminosity and effective temperature that
develop shortly after model $1490$. The star passes through a local
luminosity minimum after the 6th pulse cycle. Around this local
luminosity minimum, the surface-luminosity variation reaches its
maximum with a relative amplitude of $5$~\%. During the ensuing upward
evolution along the lower AGB, the amplitude of the ThMPs decreases
continuously (see also Fig.~\ref{fig:m30dpulses}). The variability
disappears completely at about the epoch labeled by E (model number
$3234$) when the helium in the center has essentially burned away,
i.e. having reached $Y_{\mathrm{c}} = 1.15 \cdot 10^{-3}$.

The temporal evolution of pertinent luminosity components is shown in
Fig.~\ref{fig:m30dpulses}.  At model $1490$, the epoch of the onset of
the ThMPs, the star has reached an age of about $395$~My. The
amplitude of the luminosity variation in the He-burning shell
($L_{\mathrm{He}}$) grows rapidly to reach maximum
amplitude~--~varying between $10$ and $110 \lsol$~--~in cycle
$7$. Since the layers in and above the He shell expand during the
increase of the shell's energy production, the hydrogen shell
effectively cools so that its luminosity, $L_{\mathrm{CNO}}$, drops.
Figure~\ref{fig:m30dpulses} shows also the mirror-image behavior of
the mechanical work, $L_{\mathrm{g}}$, in the star's gravitational
field. Energy is required for expansion during the $L_{\mathrm{He}}$
rise and it is gained during the shrinking of the envelope around
$L_{\mathrm{He}}$ minimum. In the sum of the partial luminosities,
$L_{\mathrm{tot}} \equiv L_{\mathrm{He}} + L_{\mathrm{CNO}} +
L_{\mathrm{g}}$, the total luminosity variation remained always very
small, less than $5$~\%. It is mainly $L_{\mathrm{g}}$ which
counteracts the strong variation of $L_{\mathrm{He}}$, despite the
fact that hydrogen burning mostly dominates the energy production.

\begin{figure}
      \resizebox{.98\hsize}{!} {\includegraphics{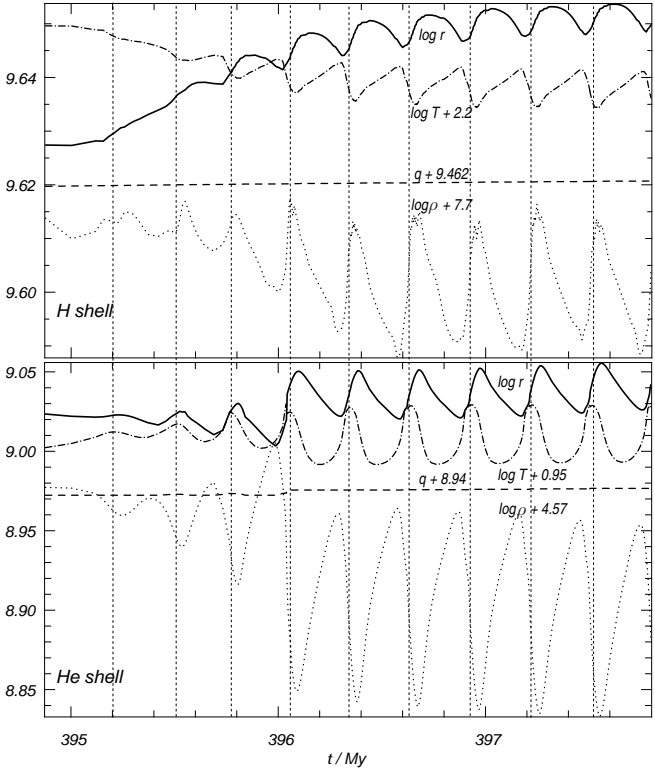}} 
\caption{
   Temporal variation of selected physical quantities at the
   center of the hydrogen-burning shell (top panel) and the
   helium-burning shell (bottom panel). The vertical dashed lines
   indicate the epochs of the first nine $L_{\mathrm{He}}$ peaks of
   the ThMP episode. The loci of temperature ($T$), fractional mass
   ($q$), and density ($\rho$) were shifted vertically to fit the
   figures that were set up to display the variation of the radii ($r$) 
   of the shell centers.
} 
\label{fig:thmpshells}
\end{figure}
Figure~3 shows, for the H-burning shell (top panel) and the
He-burning shell (bottom panel), the temporal evolution of physical
quantities at the center of the corresponding nuclear-burning shell
over the first nine ThMPs. The data for temperature, $T$, density,
$\rho$, and the fractional mass, $q$, were shifted vertically to
fit into a decently scaled single diagram. The vertical dashed lines
in Fig.~3 mark the epochs of maximum $L_{\mathrm{He}}$. The mean radii
of the H- and He-shell slowly increase with time reflecting the slow
ascent of the stellar model along the second giant branch
(cf. Figure~\ref{fig:m30tracks}). Notice the quick saturation of the
amplitude growth of the radius variation of the He shell after about
the first four pulse cycles.  The increase in $q$, the fractional
mass, of both shells is small. In the He-shell, the density minimum
occurs at minimum radius of the shell and vice versa; maximum
temperature is well in phase with $L_{\mathrm{He}}$ reflecting the
dominating dependence on temperature of the nuclear energy-generation
rate of helium burning.  The H-shell behaves differently: Minimum
density agrees with minimum radius and with maximum temperature (this
coincides with the maximum-density phase of the He-shell). Hence,
minimum radii of H- and He-shell coincide whereas maximum radius of
the H-shell is reached slightly later than maximum radius of the
He-shell. The wiggling in $\log\rho$ at the center of the H-shell is
an artifact contracted from our definition of the shell center as the
gridpoint with maximum nuclear energy generation rate; this maximum
slightly shifts in mass and as the density contrast across the H-shell
is significant, numerical noise is easily picked up.  

After a total of $30$~pulse cycles, at about $403.7$~My (model number
$3234$), the ThMPs die out. Hence, the ThMP episode lasts for about
$8.7$~My. The mean pulse-cycle length ($\tau_{\mathrm{ThMP}}$) is
$0.29$~My, a value that remains very stable throughout the whole
instability episode.

\subsection{The secular stability analyses}
\label{sect:secstab}
\begin{figure*}
\sidecaption
 \includegraphics[width=11cm]{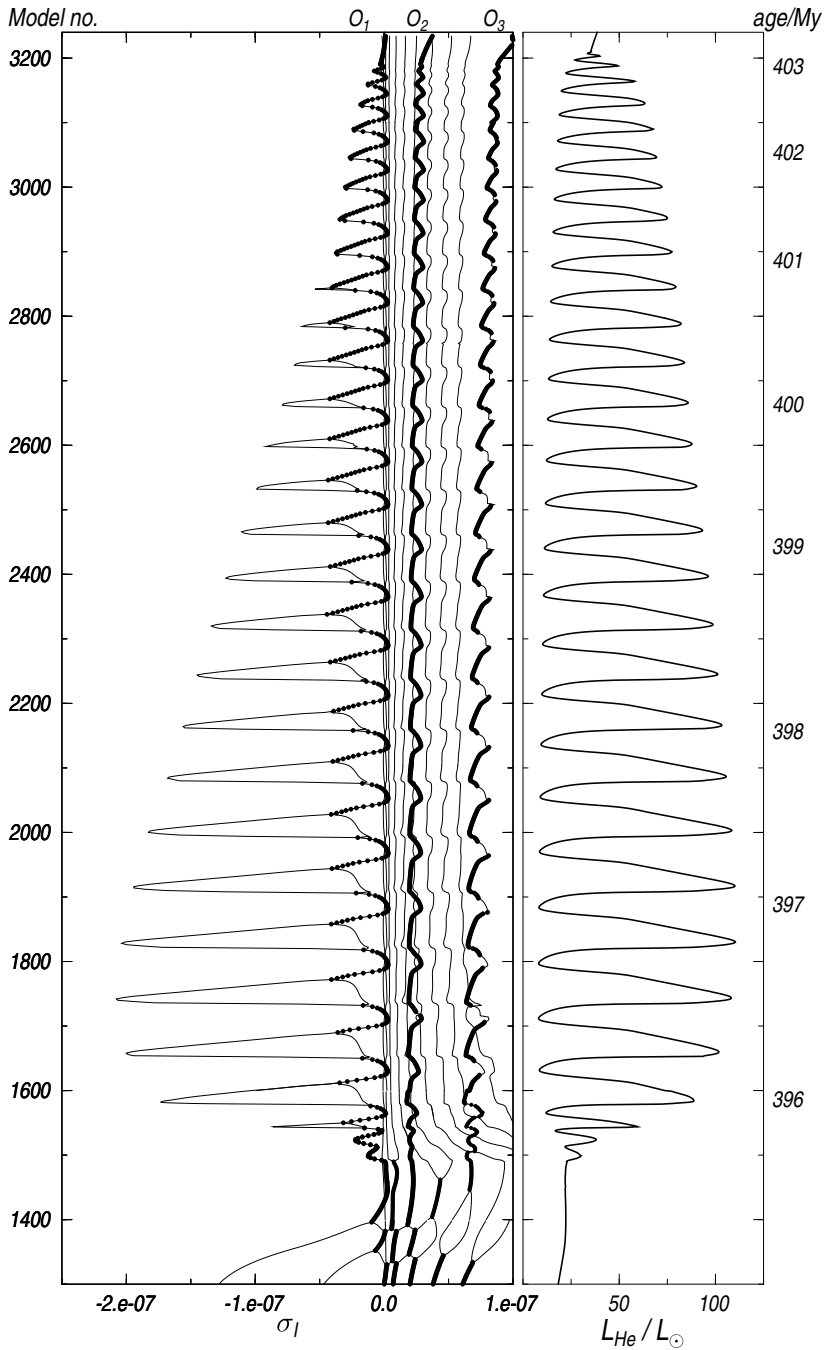} \caption{The left panel of
      the graph shows the modal diagram with the imaginary parts ($\sigi$)
      of the secular modes of the model series presented in
      Fig.~\ref{fig:m30tracks}.  The evolutionary epochs defining the
      ordinate are counted in model numbers on the left and in
      mega-years on the right-hand side. Oscillatory secular modes,
      with non-vanishing real parts of the eigenfrequencies, are
      displayed as filled circles.  For reference in the text, the
      three main branches of oscillatory secular modes extending over
      most of the ThMP phase are labeled by O$_1$, O$_2$, and O$_3$.
      To ease the association of model numbers with a characteristic
      physical quantity of that evolutionary epoch, the panel on the
      right displays the temporal evolution of the luminosity
      generated by helium burning.}
\label{fig:modiagim30}
\end{figure*}

Between model numbers $1300$ and $3234$, i.e. between labels A and E
of Fig.~\ref{fig:m30tracks}, we performed linear stability
computations on the evolutionary models to study their secular
eigenspectra.  The physical quantities were assumed to change
proportional to $\exp(\imag \omega t)$.  
The quantity $\sigma$ that is used in the following is the frequency
$\omega$ divided by $\sqrt{3 G \mast / \rast^3}$ with $G$ being the
gravitational constant.  To obtain approximate e-folding times in
mega-years in Fig.~\ref{fig:modiagim30}, compute $5.12\times
10^{-9}/\sigi$.  

For the stability computations we relaxed the requirement of thermal
equilibrium of the stellar models; the assumption of complete
equilibrium is what is usually implemented in stellar pulsation
analyses. For the secular problem, in particular when pulses develop,
$L_{\mathrm{g}}$ contributions might be significant so that refraining
from the assumption of thermal equilibrium appears more
appropriate. The linear stability equations, containing approximations
for the thermal imbalance terms, are given in Appendix A.

The left panel of Fig.~\ref{fig:modiagim30} shows $\sigi$, the
imaginary parts, of the lowest few secular eigenmodes as a function of
model number. Complex conjugate eigensolutions (oscillatory secular
modes) are indicated by full dots. Dot-free lines indicate vanishing
real-parts of $\sigma$, i.e. purely monotonic modes. For better
correlation of the linear results with the evolutionary computations,
the right-hand panel of Fig.~\ref{fig:modiagim30} plots the
$L_{\mathrm{He}}$ oscillations.  For completeness, the age in
mega-years is labeled along the rightmost ordinate of the figure.

The complexity of the secular eigenspectrum is considerable; and this
is so already before the onset of the ThMPs. During the late core
helium-burning phase, the modal diagram is a complicated web of
complex-conjugate modes unfolding into pairs of non-oscillatory modes
that later merge again into complex-conjugate solutions in different
arrangements. In all circumstances, the number of modes involved is
conserved.

At model $1300$, two unstable monotonic secular modes were
encountered. As the core dominance of the nuclear burning ceased, the
strength of the instability diminished. Eventually, the most unstable
secular modes merged with other branches into oscillatory modes. Most
of them became or remained stable during the early helium
shell-burning phase. A discussion of the secular-stability behavior
during the core He-burning phase of intermediate-mass stars is
postponed to a forthcoming paper.

The onset of the ThMP phase manifested itself in the least damped
oscillatory secular mode going overstable. The first two ThMP cycles
appeared as unstable oscillatory secular modes in the linear
analysis. Beginning with cycle three, the unstable complex conjugate
eigensolution unfolded into a pair of monotonic modes during the
maximum of $L_{\mathrm{He}}$ within a cycle.  During phases of reduced
efficiency of the helium burning shell, the two monotonic modes merged
to go into the next cycle through a short damped episode. The cyclic
unfolding of the unstable oscillatory secular mode continued to cycle
$21$ after which the amplitude of the ThMP was seemingly small enough
for the unstable secular mode to remain purely oscillatory. The ThMPs,
as seen in $L_{\mathrm{He}}$ (e.g. right panel of
Fig.~\ref{fig:modiagim30}), terminate by the dominant unstable oscillatory
secular mode going stable but remaining oscillatory.

The details of the behavior of the complex eigenvalue $\sigma$
throughout the phases of the ThMP episode must not be
over-interpreted. After all, the stability analyses were performed on
the thermally pulsing models themselves.
\footnote{
In analogy with the pulsation theory of stars, our linear secular
analyses can be compared to performing linear pulsation analyses on
nonlinearly pulsating star models. The period of pulsation derived
from such a linear analysis varies cyclically proportional to
$R_\ast(t)^{3/2}$ where $R_\ast(t)$ is the nonlinearly pulsating
stellar radius. Also the damping/excitation rate (our $\sigi$ here)
varies along a pulsation cycle. In particular when the nonlinearly
pulsating model is hottest, the pulsational excitation by the
$\varkappa$-mechanism of the stellar envelope is diminished as the 
driving zone lies too close to the surface; at the phase of lowest
surface temperature, excitation is diminished as the driving region
lies too deep in the envelope.}
Still, to gain access to the conditions under which complex secular
eigenvalues unfold into real pairs, secular analyses on the full
cycles of ThMP models might prove helpful.

In addition to the unstable, mainly oscillatory secular mode,
Fig.~\ref{fig:modiagim30} shows a damped oscillatory mode, O$_2$ that
never unfolded during the whole ThMP episode.  The oscillatory secular
mode baptized O$_3$, on the other hand, unfolded again cyclically into
pairs of monotonic modes, but in this case during the minimum phases
of $L_{\mathrm{He}}$; Fig.~\ref{fig:modiagim30} shows only one of the
monotonic branches of O$_3$ since a follow-up of both of them turned
out to be computationally cumbersome due to other close-lying secular
modes.

In the range of the $\sigi$ domain considered here, we encountered
only three oscillatory or partially oscillatory secular modes. At
least in the $\sigi$~--~model number (or age) diagram these
oscillatory modes overlap with purely monotonic secular modes. Except
for the mode with the longest time-scale, all of the monotonic modes
are \emph{always} damped.  Note that the crossing of modes in
Fig.~\ref{fig:modiagim30} is a projection effect due to plotting
$\sigi$ only instead of the complex $\sigma$; on the complex $\sigma$
plane, the modes are well separated.

The secular mode with the smallest $\sigi$ values and which remains
always monotonic switches between weakly stable and unstable; the
growth rates of this monotonic mode are two orders of magnitude
smaller than those of the overstable complex secular mode. After model
$3216$, when the oscillatory secular mode went stable and ThMPs
disappeared, the lowest monotonic secular mode remained unstable
with an e-folding time of $3.4$~My (as a comparison, an evolutionary
change of the total luminosity by 5\% takes $3.1$~My at this epoch of
evolution).  

For the set of secular modes that remain monotonic at all epochs after
the onset of the ThMPs, the $\sigi$'s vary all in phase during a pulse
cycle.  Furthermore, the general $\sigi$ evolution during the ThMP
episode is the same for all these monotonic secular modes; and this
evolution differs from the oscillatory secular modes. The latter fact
is most evident in the evolution of $\sigi$ of the most strongly
damped oscillatory mode relative to the monotonic secular modes
between models $1500$ and about $1800$.

The stellar model sequence whose computation included diffusive
overshooting and which showed no evidence at all of thermal pulses in
the evolution computations also revealed \emph{no} overstable
oscillatory secular mode at any time during the double-shell nuclear
burning phase. However, from the very beginning of the core helium
burning through the double-shell burning phase at least one monotonic
secular mode was always unstable.  Overshooting left its imprints in
the abundance structure and in a bigger stellar core as compared with
ThMP models.

\subsection{On the suppression of thermal micropulses}
\label{sect:suppressedpulses}
\begin{figure}
      \resizebox{.98\hsize}{!}
      {\includegraphics{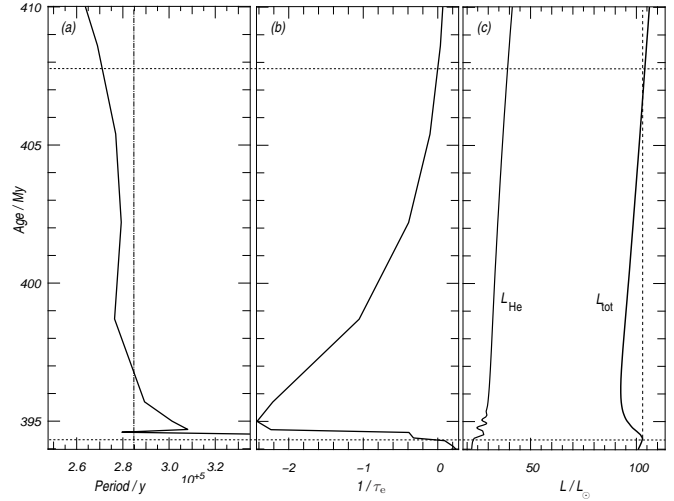}} \caption{ Panels (a)
      and (b) show eigendata of the overstable oscillatory secular
      mode as computed in the pulse-suppressing model sequence
      which is characterized in panel (c) by the helium-luminosity
      ($L_{\rm He}$) and the total luminosity ($L_{\rm tot}$)
      evolution. Panel (a) displays the period evolution of the
      overstable oscillatory secular mode as a function of model
      age. The vertical line shows the magnitude of the average
      pulse-cycle length that was deduced from
      Fig.~\ref{fig:m30dpulses}. The inverse of the e-folding time
      $\tau_{\rm e}$ (measured in units of mega-years) of the
      overstable secular mode is plotted in panel (b).}
\label{fig:supprsspulses}
\end{figure}

The ThMPs are found to be a delicate phenomenon. First of all, ThMPs
are encountered only in computations with appropriate convection
treatment and in the narrow mass range between about $2.5$ and about
$3.5 \msol$. The easiest and in some sense the most dangerous way to
miss the ThMPs is by means of the time-step choice to evolve a model
star. Already \citet{schwarm65} cautioned in their paper that
inappropriately large time-steps, of the order of twice the e-folding
time of the instability, suffice to numerically suppress a pulse. We
repeated the computation described in Sect.~\ref{sect:thmps_evol} (we
will refer to it to as the ``original'' one) by enforcing a larger
time-step. The coarser-grained time evolution sequence started helium
shell burning with a time-step that was $1.88$ times larger than the
one of the original sequence. The structure (position and thickness)
of the helium shell and the abundance profiles differed only
marginally from those in the original computation. Nonetheless, the
helium shell remained essentially stable: We only observed a short
initial oscillatory phase of $L_{\mathrm{He}}$ lasting for roughly
three strongly damped cycles. The first cycle had the largest
amplitude with a relative $L_{\mathrm{He}}$ variation of
$11$~\%. (Remember, in the original evolution computation, the
relative ThMP amplitude of the helium shell reached $400$~\%.) No
variability was manifested in $L_{\mathrm{tot}}$. Even an inspection
at highest numerical resolution of the computed observable stellar
parameters revealed no sign of an instability whatsoever (see panel c
of Fig.~\ref{fig:supprsspulses}).
  

Figure~\ref{fig:supprsspulses} displays results from computations
performed on the pulse-suppressing stellar evolution sequence. As
mentioned before, we found one overstable secular mode with
characteristics that were comparable to the overstable secular mode in
the thermally pulsing model sequence. The dashed horizontal lines
extending across all three subpanels of the figure indicate, at the
bottom, the onset of the secular instability and the termination of
the instability in the upper part of the plot.  Panel (a) shows the
period evolution of the overstable secular mode as a function of model
age; the dash-dotted line in this panel indicates the mean pulse
period as derived from the fully pulsing evolutionary sequence. Note
that the period obtained in the linear analysis deviates by less than
10\% from the completely independently computed nonlinear cycle length
of the pulses. The inverse of the e-folding time (measured in
mega-years) of the instability is plotted in panel (b). For most of
the instability phase, the e-folding time, $\tau_{\rm e}$, is less
than or of the order of the evolutionary time-scale itself. Panel (c)
finally shows the luminosity evolution of helium burning ($L_{\rm
He}$) and of the star's total luminosity, $L_{\rm tot}$. The quantity
$L_{\rm He}$ of the evolution computation shows only a weak sign of
instability that lasts for roughly three strongly damped cycles; the
total luminosity carries even no sign at all of an instability. We
emphasize again that the phase of instability as obtained from the
linear secular analysis is comparable to the evolutionary phase over
which the ThMPs are encountered in the ``non-suppressing'' model
sequence; this is traced out with the vertical line plotted from the
local maximum of $L_{\rm tot}$ in panel (c). The local maximum at the
bottom of the figure coincides with the epoch of model 1490 shown in
panel (c) of Fig.~\ref{fig:m30tracks}. In the ``non-suppressing''
model sequence, the ThMPs are superimposed on the ensuing broad
$L_{\rm tot}$ trough; the luminosity of the terminal model E (whose
oscillatory secular modes were all at least weakly stable) lied
slightly below the total luminosity of model 1490. In the
pulse-suppressing sequence, on the other hand, the linear instability
extends to roughly the epoch when the luminosity reaches again the
level of the initial bump. 

\section{Discussion}
\label{sect:discuss}
In accordance with \citet{haerschild72} we found the secular
eigenspectrum to show an oscillatory secular mode going unstable at
the epoch of the ThMPs setting in. The secular mode O$_1$ remained purely
oscillatory through the first two cycles. Starting with the third
cycle, O$_1$ unfolded temporarily into a pair of purely monotonic
modes, both of which were unstable.  The monotonic-mode interlude took
up roughly $46$~\% of a cycle (at cycle $7$) with a slow decrease of
this percentage as the ThMP episode progressed. After cycle $21$,
O$_1$ remained again purely oscillatory up to the last ThMP cycle
(number $30$). It was always around the $L_{\mathrm{He}}$-maximum
phases where the oscillatory O$_1$ mode unfolded. Interestingly
enough, O$_3$, the other oscillatory secular mode with eigenvalue
unfoldings, showed these to occur at around the phases of minimum
$L_{\mathrm{He}}$. 
\begin{figure}
      \resizebox{.85\hsize}{!} {\includegraphics{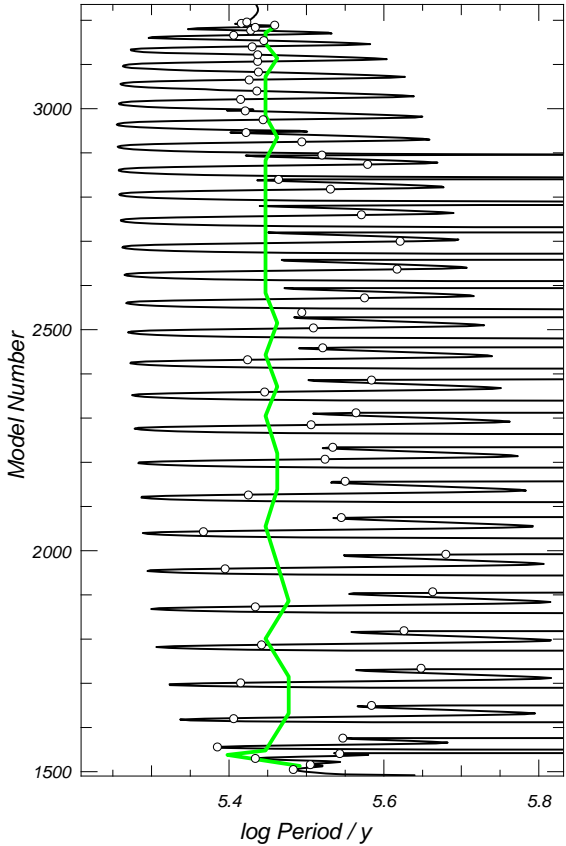}}
      \caption{ The periods of the overstable oscillatory secular mode
      O$_1$ shown in Fig.~\ref{fig:modiagim30} during the ThMP episode
      as a function of model number. The almost horizontally running
      lines result from the unfolding and merging of secular modes. At
      the corresponding phases, the periods go formally to
      infinity. The thick grey line indicates the cycle lengths that were
      derived directly from the $L_{\rm He}$ variation of the
      evolutionary computations. The open circles indicate the periods
      as measured at pulse phases with $L_g = 0$, i.e. at
      `equilibrium' epochs.}
\label{fig:periods30d}
\end{figure}

Figure~\ref{fig:periods30d} shows the periods computed for the mode
branch O$_1$ as black continuous line. The lengths of the ThMP cycles
derived directly from the evolutionary computations are plotted as a
grey line. The sharp horizontal excursions in the figure are caused by
the unfoldings of the complex conjugate eigenvalues and by the merging
of two monotonic secular modes where the period goes formally to
infinity.  The agreement between linear theory and the cycle lengths
deduced from the nonlinear stellar evolution is reasonable.  The
variation of the ``linear'' period derived from the secular stability
analysis during a ThMP cycle is reminiscent of the variation of the
pulsation period computed on a nonlinearly pulsating star at various
phases of a pulsation cycle (cf. footnote in Sect~\ref{sect:secstab}).
The open circles denote the periods as measured at phases with $L_g =
0$, i.e. at formal equilibrium phases. The maximum deviations of the
cycle lengths derived from the evolutionary computations and the
periods at equilibrium phases do not exceed $40\%$.  The
correspondence of the two differently computed cycle lengths in
Fig.~\ref{fig:periods30d} are considered as a good indication that a
linear secular analysis is appropriate to at least qualitatively
understand the cyclic variability of the ThMP phase; this point of
view is further supported by the results of the secular analyses of
the pulse-suppressing model sequence which was presented in
Sect.~\ref{sect:suppressedpulses}. The periods obtained in the
pulse-suppressed series agree very well with the cycle lengths
measured in the evolution computations and they are well bracketed by
the secular analyses of the thermally pulsing models.

The rest of the discussion section is structured as follows: First, we
comment on the local criteria for secular instability and the
difficulty of their use in case of oscillatory secular modes. A short
digression on the attempts to understand what discriminates between
monotonic and oscillatory secular modes follows. After that, we delve
into the properties of the linear eigenmode spectrum as we computed
it. We close the discussion section with checking the quality of the
Gabriel-expansion during the short-term evolution of the ThMP phase.

\subsection{Local stability criteria }
\label{sect:localcriteria}
The paper of \citet{schwarm65} contains a simple conceptual model
of a nuclear-burning shell in a star. The formulae they obtained at the
end of their local analysis allow to quickly estimate the shell's
stability against thermal perturbations; therefore, the local criteria
were used ever since in computational research and in textbooks.  It
turns out that two conditions must necessarily be fulfilled for
\emph{instability}: The shell must be thinner than
\begin{equation}
\label{eq:lclshellstab} 
\frac{\Delta r}{r} < \frac{\Gamma_1}{4}\cdot f \,,
\quad\mathrm{but~sufficiently~thick~so~that}\quad
\frac{\Delta T}{T} > \frac{4}{\epst} \,,
\end{equation}    
the quantity $f$ measures the deviation from a homologous
perturbation; most frequently $f$ is chosen ad hoc to be about unity,
i.e. close to homology.  The variation of physical quantities across
the nuclear burning shell are indicated with a leading $\Delta$. The
width of the shell is not uniquely defined and gives rise to
considerable ambiguity.  The remaining quantities take on appropriate
mean values that are representative of the nuclear-burning shell.
With $\epst$ we denote $\partial \log \varepsilon / \partial \log T$
at constant pressure.  The rest of the quantities have their canonical
meanings.  
Note that the instability that is described with conditions
(\ref{eq:lclshellstab}) is a monotonic one; this simple
one-zone~--~type model does not admit of complex solutions at
all. Therefore, relying on criteria (\ref{eq:lclshellstab}) in the
case of thermal-pulse cycles that start via oscillatory secular modes
can be misleading.

Table~\ref{table:locanal} contains numerical values of the two sides
of both instability criteria in (\ref{eq:lclshellstab}) for selected
stellar models and for various choices of the width of the He-burning
shell. The thickness of the shell was measured as the extension over
which nuclear energy generation exceeds a prescribed fraction of
$\varepsilon_{\rm max}$.  The range of numbers given in column 
``$4/\epst$'' is caused by the non-negligible variation of $\epst$
across the nuclear burning shell. Model $1470$ was secularly stable,
it showed no unstable secular modes and around that epoch $L_{\rm He}$
varied still only weakly and monotonously.  At model number $1516$,
$L_{\rm He}$ just passed through the first peak and mode O$_1$ was
overstable at this epoch; the ThMP-phase just developed. Finally,
model number $3234$ lies after the ThMP phase with no overstable
oscillatory secular modes anymore and $L_{\rm He}$ changed again
slowly and monotonously.

\begin{table}
\caption{Evaluation of physical quantities across the He-shell}
\label{table:locanal}                 
\centering                            
\begin{tabular}{c c c c c c}          
\hline\hline                          
Model & width & $\Delta r / r$ & $\Gamma_1 / 4$ 
              & $\Delta T / T$ & $ 4 / \epst  $ \\  
\hline                                
1470 &$\varepsilon_{\rm max}/10$  & 0.122 & 0.41 & 0.067 & $0.1 - 0.09$ \\ 
     &$\varepsilon_{\rm max}/100$ & 0.23  & 0.41 & 0.129 & $0.1 - 0.09$ \\ 
\smallskip
     &$\varepsilon_{\rm max}/1000$& 0.318 & 0.41 & 0.168 & $0.1 - 0.08$
     \\ 
1516 &$\varepsilon_{\rm max}/10$  & 0.175 & 0.41 & 0.067 & $0.11 - 0.09$ \\ 
     &$\varepsilon_{\rm max}/100$ & 0.258 & 0.41 & 0.109 & $0.11 - 0.09$ \\ 
\smallskip
     &$\varepsilon_{\rm max}/1000$& 0.349 & 0.41 & 0.149 & $0.11 - 0.08$
     \\ 
3234 &$\varepsilon_{\rm max}/10$  & 0.413 & 0.41 & 0.149 & $0.32 - 0.11$ \\ 
     &$\varepsilon_{\rm max}/100$ & 0.514 & 0.41 & 0.206 & $0.32 - 0.13$ \\ 
     &$\varepsilon_{\rm max}/1000$& 0.585 & 0.41 & 0.224 & $0.32 - 0.08$
     \\ 
\hline                                   
\end{tabular}
\end{table}

Most frequently, only the first condition of the instability criteria
(\ref{eq:lclshellstab}) is referred to in the literature; i.e. only
the test if a shell is sufficiently thin for the instability to
develop.  The instability condition in expressions
(\ref{eq:lclshellstab}) is equivalent to the physical notion of a
generalized specific heat (also referred to as ``gravothermal''
specific heat)
\footnote{To explain the physical processes during a secular
instability, \citet{kippenhahn70} was the first to refer to an
``effective specific heat'' that goes negative. The baptizing of this
effective specific heat as ``gravothermal specific heat'' seems to go
back to \citet{sugimiya81} who were guided by an analogy with the
gravothermal catastrophe of self-gravitating systems.}
which goes positive within the nuclear burning
shell. \citet{giannoneweigert67} generalized the local instability
criteria (\ref{eq:lclshellstab}), adopting a more general equation
of state and allowing for more variation of the physical background
quantities across the nuclear-burning shell.
\citet{yoonlangersluys04} elaborated on the local stability criteria
of nuclear burning shells; building upon the approach of
\citet{giannoneweigert67}
\footnote{The \citet{giannoneweigert67} derivation of the local stability
criteria of nuclear burning shells is more general than what
Schwarzschild \& H\"arm (1965) (SH65) did. Not only do Giannone \& Weigert
accommodate a more general equation of state than SH65, they are also not
resorting a priori to homologous perturbations. The resulting
stability criteria look considerably more crowded; on the other hand
one gains a broader range of applicability than the SH65 one. This is a
point of interest when objects other than low-mass AGB stars are studied.
} 
but still with the simplifying assumption of homologous
perturbations. The aim was to quantitatively test stellar evolution
models towards thermal-pulse instabilities.  The problem with the
local stability criteria lies mainly in the necessity to have to
assume a functional form of the perturbations (usually homologous
motion) and therefore to constrain the nature of the instability and
to have to adopt a not well definable thickness of the shell.
\citet{yoonlangersluys04} mention that the shell thickness in their
helium-accreting white dwarf changes by about $15 \%$ if they use once
the radii at $r(\varepsilon = 10^{-2} \varepsilon_{\mathrm{max}})$ and
once at $r(\varepsilon = 10^{-3} \varepsilon_{\mathrm{max}})$ with
$\varepsilon_{\mathrm{max}}$ being the peak nuclear burning rate. For
their discussions, Yoon et al. chose $r(\varepsilon = 2\times
10^{-3}\cdot \varepsilon_{\mathrm{max}})$ as then their local analyses
agreed best with the evolutionary computations. This approach might be
incomplete, however, since, as we have discussed before, the
dissipation inherent in the computational schemes for stellar
evolution might suppress thermal pulses despite their being physically
present and detectable in a full linear secular analysis.


In Table~\ref{table:locanal}, local secular instability obtains if the
numbers of column ``$\Delta r / r$'' are smaller than those of column
``$\Gamma_1/4$'' and if the numbers of column ``$\Delta T/T$'' are
bigger than those of column ``$4/\epst$''. For model 1470, the
geometrical criterion in (\ref{eq:lclshellstab}) indicates instability
for all choices of shell thickness; the thermal part of the local
condition favors instability for sufficiently thick shells (i.e. for
$\varepsilon_{\mathrm{max}}/100$ and
$\varepsilon_{\mathrm{max}}/1000$). Note that for model 1470
(cf. Fig.~\ref{fig:modiagim30}) neither the solution of the full
secular boundary-value problem nor stellar evolution shows a sign of
secular instability yet. For model 1516, the geometric and the thermal
local criteria behave the same as for model 1470. In this case, it is
compatible with stellar evolution and full secular-analysis
results. For model 3234, at an epoch in the post-ThMP phase, all
choices of shell size point to stability and, depending on the choice
of a representative magnitude of $4/\epst$, the same applies to the
thermal criterion.

In agreement with \citet{yoonlangersluys04} we find that the
particular choice of shell thickness influences crucially the result
of the local criterion.  Moreover, the onset of the ThMPs is not
captured reliably with the local stability criteria. This does not
further surprise as the ThMPs are the result of the instability of a
complex secular mode and the local criteria are based on a one-zone
model description that do not support complex solutions at all.  
 
In the thermal criterion of conditions (\ref{eq:lclshellstab}),
thickness is expressed as temperature and luminosity differences
across the burning shell, these quantities are more sensitive to a
particular choice of shell-boundaries than the radius measures;
therefore the thermal criterion is more sensitive to a particular
choice of shell thickness.  

\subsection{Oscillatory versus monotonic secular solutions}
\label{sect:osc_vs_mon}
Already \citet{GabrielNoels72} stressed  that due to the
non-hermiticity of the secular stability problem, complex eigenvalues
must not be considered unusual. An unambiguous prescription of
conditions that must prevail in a star to lead to complex
eigensolutions have not been put forth in the literature so far. 

As mentioned at the beginning of Sect.~\ref{sect:discuss}, the
merging/unfolding phases of the O$_1$ mode branch during the pulse
cycles differ from that of O$_3$ (see Fig.~\ref{fig:modiagim30});
this immediately enforces the important conclusion that not stellar
background quantities alone discriminate between monotonic and complex
secular eigensolutions, an aspect already conjectured by
\citet{defouw73}. Despite adopting the point of view that the thermal
pulses themselves are (nonlinear) relaxation oscillations,
\citet{defouw73} emphasized the combination of hydrostatic readjustment
to a heat perturbation and the magnitude of the temperature dependence
of nuclear burning, $\epst$, to decide over the nature of the secular
eigenvalue. Earlier on, \citet{dennis71} studied the dependence on
$\epst$ of the secular stability of the helium-burning shell of
massive stars. Low values of $\epst$, left the considered secular mode
oscillatory damped; increasing $\epst$ drove it overstable and for
$\epst > 53$, it eventually unfolded into two monotonic branches. In
our computations, we found the magnitude of $\epst$ at the peak of the
unstable He-burning shell to stay at about $40$ and to vary by less
than $5\%$ during the ThMP cycles; in the case of the unfolding
and re-merging of the O$_1$ mode, this variation was furthermore
opposite to what \citet{dennis71} suggested for this process.  Hence,
together with the phase-shifted merging/unfolding behavior of O$_3$,
we conclude that the importance of $\epst$ lies at best in its
magnitude but not in its variation. 


\subsection{On the secular-eigenmode spectrum}
\label{sect:eigenspectrum}

Looking at the modal diagram in Fig.~ \ref{fig:modiagim30}, we
realize that the oscillatory and the monotonic secular modes evolve
differently on the $\sigi$~--~age plane . This applies also to
cyclically repeating episodes of unfolding of complex eigenmodes into
two monotonic modes during the ThMP phase and in particular it is true
for the early ThMP phase when the He shell source takes over from
central helium burning. However, throughout the whole ThMP phase the
stable oscillatory secular modes (O$_2$ and O$_3$) show an average
$\sigi$ increase that exceeds that of the monotonic modes; this
suggests that oscillatory and monotonic secular modes are each
influenced by different regions of the stellar interior.  
Unfortunately, the secular eigenproblem admits of no special
mathematical structure, so that all the helpful tools developed for
the hermitian adiabatic pulsation problems do not grip. We must
therefore resort to analyze eigenfunctions at selected instances and
hope to derive therefrom useful hints on the nature of the secular
modes.

In the following, we refer to the eigenfunction components $\zeta$,
$p$, $t$ and $\delta q \equiv T\,\delta s$ standing for the relative
radius, pressure, temperature, and heat perturbation, respectively. We
studied in more detail the eigenfunctions at the begin of the ThMP
episode: For model $1488$, just before the onset of the thermal
micropulses, for model $1492$, one stellar model after the onset of
the micropulses, and for model $1516$, at the $L_g = 0$ phase after
the first pulse cycle.
\begin{figure}
      \resizebox{.97\hsize}{!}{\includegraphics{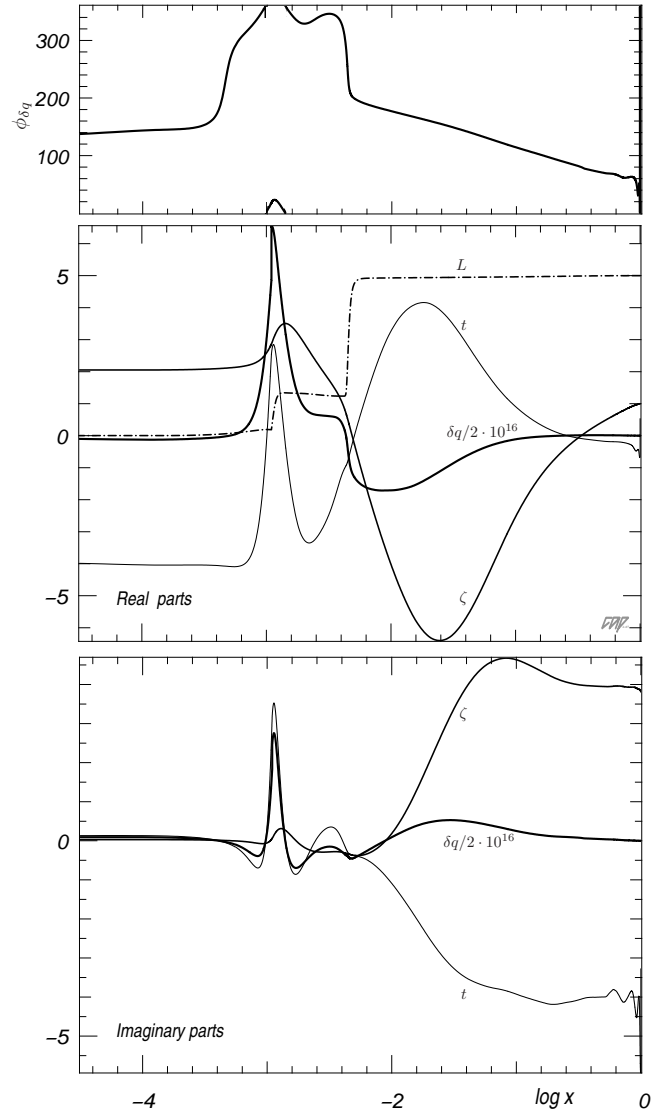}}
\caption{ Spatial run, measured in
      fractional radius $x$, of eigenfunction data of the O$_1$ mode
      in model $1516$. The two lower panels show the real and
      imaginary parts of the relative temperature perturbation $t$,
      the radial displacement $\zeta$, and the heat perturbation
      $\delta q$. The heat perturbation and the luminosity $L$
      (dash-dotted line in the middle panel) of the stellar model are
      suitably scaled to fit the plot range. The top panel shows the
      spatial variation of the complex phase angle of the heat
      perturbation $\delta q$.}  
\label{fig:efso1_1516} 
\end{figure}
Since, as mentioned before, the secular problem admits of no favorable
mathematical structure, we cannot assume any ordering of the modes
(e.g. no correlation of node numbers and the magnitude of the
eigenfrequency) and indeed this is what we encountered. The counting
was exemplarily done on eigenfunctions of the lowest five monotonic
eigenmodes, and the first three oscillatory ones (O$_1$, O$_2$,
O$_3$). Node counting revealed that the mechanical eigenfunction
components and the thermal ones behave differently. For all modes
considered, the mechanical eigenfunction components had either $2$ or
$3$ nodes; i.e. the number of nodes of the displacement and the
relative pressure perturbation was independent of the particular mode.
Furthermore, the number of nodes of a chosen mode was subject to
change (by one or two nodes) as the star model evolved. For all
secular modes, monotonic and oscillatory, the radial displacement
admits a node either within or close to the hydrogen burning shell.

Figure~\ref{fig:efso1_1516} plots representative eigenfunctions
of mode O$_1$ in model $1516$. The top panel shows the spatial
variation of the phase of the heat perturbation $\delta q$. The middle
panel displays the real parts of $\zeta$, $t$, and of the suitably
scaled $\delta q$; the corresponding imaginary parts of the
eigenfunctions are contained in the bottom panel. For better
orientation, the middle panel is complemented with a dash-dotted line
tracing the (again suitably scaled) spatial variation of the star's
luminosity. The node of $\zeta $ at the H-burning shell is reminiscent
of the ``mirror principle'' encountered in the spatial variation of
the radius evolution of many shell-burning stars. Only mode O$_1$ has
a relatively large displacement amplitude also interior to the helium
burning shell. The displacements of modes O$_2$ and O$_3$ become
marginal already below the hydrogen-burning shell. 
On the other hand, many of the monotonic modes which we studied
have non-negligible displacement amplitudes in the deep interior with
these amplitudes growing with $\vert \sigi\vert$. Also the monotonic
secular modes admit of a node close to the H-burning shell. The
outermost node lies in the envelope with its position in radius to
grow with $\vert \sigi\vert$, for all these monotonic modes, the
maximum amplitude is reached at the surface. 
For \emph{all modes} considered, the perturbations become homologous below
$\log x \approx -3.5$; i.e. homologous variation prevails only
over a tiny fraction of the star's radius and in particular it does
not obtain at and above the nuclear burning shells. Focusing on the
perturbations of thermal quantities revealed that e.g. $\delta q$ of
the oscillatory secular modes (O$_1$ to O$_3$) is considerably
confined to the intershell region of the stellar models with some
noticeable amplitude at the base of the star's envelope, just above
the hydrogen shell. In contrast, the heat perturbations of the
monotonic modes achieve their largest amplitudes in the very deep
stellar interior with the heat perturbations in the nuclear-burning
shell region being relatively small. Based on the behavior of the
thermal perturbations, we can consider the oscillatory modes as
functionally different from the monotonic ones. Furthermore, the
thermal eigenfunctions show a node behavior that differs from that of
the mechanical ones. The number of nodes of the thermal components
exceed that of the mechanical ones in all models considered. In
contrast to the mechanical eigenfunctions, the thermal eigenfunctions
of different mode branches have always different numbers of nodes
(e.g. at model $1488$: $4$ nodes in $\delta q$ for O$_1$, $7$ for
O$_2$, and $11$ for O$_3$). But again, as for the mechanical
eigenfunctions, the number of nodes of the thermal eigenfunctions of a
particular mode branch is not constant as the model evolved.


It is worthwhile to point out that, in contrast to the findings of 
\citet{haerschild72}, the imaginary parts of the
eigenfunctions of the oscillatory modes are always comparable to the
real parts over most of the stellar interior; this applies for the
mechanical quantities as well as for the thermal variables
(cf. Figure~\ref{fig:efso1_1516}).  The phase of $\delta q$
varies significantly over most of the star's radius, in particular the
envelope is not varying synchronously; a fact which is quantified in
the top panel of Fig.~\ref{fig:efso1_1516}. The most rapid
phase variations occur in the H-burning and just below the
He-burning shell.  Across the helium- burning shell, the
phase stays essentially constant, so that eigenfunctions behave, at
least locally, like standing waves. And as to be expected, in the
region where the perturbations are homologous, i.e. for $\log x <
-3.5$, the phase variation levels off.

The phase variation of $\delta q$ is also relevant for another
aspect: \citet{haerschild72} postulated without further specification
that the complexification of secular eigenfrequencies relies on the
coupling of a ``reacting'' and a ``driving'' layer, with $\delta q $
being $\pi/2$ out of phase in the two layers. The reacting layer was
identified with a region just inside the He-burning shell; stellar
material above the helium shell was thought to have too low a heat
capacity to constitute a pertinent reactive layer and was therefore
discarded. But only the two-zone models (TZM) of \citet{defouw73}
eventually characterized the physical requirements for the occurrence
of complex eigenfrequencies in the secular problem. The TZM were
successful to reproduce results of previous numerical experiments that
aimed at understanding the occurrence of oscillatory secular modes.

\begin{figure}
       \resizebox{.94\hsize}{!} {\includegraphics{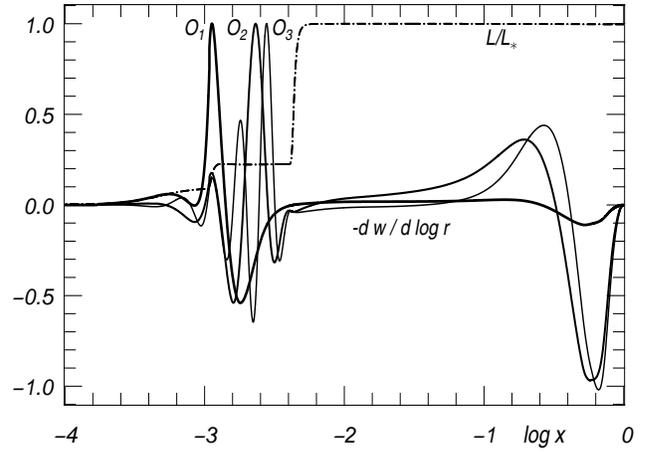}}
      \caption{ Differential work integrals $-\diff w / \diff \log r$
      throughout a star's interior. The sign convention is so as to
      show driving regions positive and damping ones negative. The
      dash-dotted line depicts, to help orientation, the spatial
      variation of the total relative luminosity, $L/L_\ast$. The data
      is shown for model $1492$, the epoch just after the onset of the
      thermal micropulses with mode O$_1$ having gone overstable.}
\label{fig:diffwork_1492}
\end{figure}

To quantify the phase difference between reacting and driving region
we resort here to the ``ultrasimplified model'' which neglects the
thermal coupling between the two zones \citep[cf. Sect. VI
of][]{defouw73}; in Defouw's notation the phase angle $\phi$ between
the heat perturbations of the TZM can be written as $\tan \phi = -
n_I/(\lambda_1 c_{11} - n_R)$. Hence, a phase difference of $\pi/2$
\citep[as referred to in][]{haerschild72} between the driving and
the reacting zone is recovered \emph{only} for the very special
circumstance of $\lambda_1 c_{11} - n_R = 0$; i.e. for $n_R=0$ and
$c_{11}=0$ which means for a neutral oscillation (temporal variation
$\propto \exp(n\cdot t)$) of a model with an infinite gravothermal
specific heat of the driving region, or for the very special
correlation $\lambda_1 c_{11} = n_R$. In all other cases, however, $0
< \phi < \pi/2$ between $\delta q$ of the driving and reacting regions,
depending on $n_R/n_I$ and $\lambda_1 c_{11}/n_I$. Together with the
fact that the heat capacity of the intershell region not being lower
than that of the He-shell, which is considered as the driving region,
we conclude that it is not unreasonable to contemplate the intershell
region or part thereof as the reacting layer. Furthermore, the
differential work plot shown in Fig.~\ref{fig:diffwork_1492} hints at a
higher reactivity of the intershell region than of the stellar
layers on the inside of the He shell.

The lower two panels of Fig.~\ref{fig:efso1_1516} display the
heat perturbation $\delta q$ as a heavy line and the relative
temperature perturbation, $t$, as a thin solid line. The displayed
spatial run is representative of the O$_1$ behavior during the whole
ThMP phase. We emphasize the positive correlation between $\delta q$
and $t$ in the He-burning shell: Locally a necessary condition for a
secular instability to develop; if a star goes eventually unstable
depends of course on the detailed damping and excitation efficiencies
throughout the whole object. (The amplitude of $\delta q$, an absolute
perturbation, was divided ad hoc by $2\cdot 10^{16}$ to match the
scale of the figure.) If $\delta q = A \cdot \delta T / T$ with $A
> 0$ lies in nuclear-burning layers with sufficiently
temperature-sensitive reactions, this helps for a marginal heat
perturbation to develop into an instability. The $\delta q = A \cdot
\delta T / T$ correlations in suitable stellar layers was encountered
only for O$_1$ modes.  The other oscillatory secular modes, despite
having also positive $\delta q$~--~$\delta T / T$ correlations, these
correlations were confined to the intershell region where no physical
agent exists to exploit the temperature perturbation at appropriate
phase. Also the monotonic modes show regions of suitable $\delta
q$~--~$\delta T / T$ correlations which were usually located in the
very deep interior, below the He-burning shell, i.e. in the inert C/O
interior.
 
To more directly illustrate the action of the helium shell to power
the instability and thermal activity of the intershell region we
computed work integrals for some overstable oscillatory secular
modes. Following the reasoning and interpretations of
\citet{glatzel94}, we justify the adoption of the work integral even
in the case of rapidly growing instabilities.

For model $1492$, Fig.~\ref{fig:diffwork_1492} displays the 
differential work curves
\begin{equation}
-\frac{\diff w}{\diff \log r} \propto
                              \mathrm{Im}\left( t^\ast p \right)
                              P\,r^3\, \delta \,,
\label{eq:diffwork}  
\end{equation}
of the oscillatory modes O$_1$ to O$_3$.  We defined $\delta \equiv
\partial \log \rho / \partial \log P$ at constant temperature. The
asterisk refers to the complex conjugate of the quantity to which it
is applied.  All differential-work curves in
Fig.~\ref{fig:diffwork_1492} are normalized independently to unity at
the positions of their respective maximum driving as we are
mainly interested to illustrate the domains of ``activity''. The
spatial variation of the differential work is representative of each
mode for the whole ThMP episode; it is only the strengths of the
various driving and damping regions that change relative to each
other. For better orientation, the thin dash-dotted line in
Fig.~\ref{fig:diffwork_1492} traces out the spatial run of the total
relative luminosity. For the O$_1$ mode, it is evident that the
maximum driving occurs in the helium-burning shell. The intershell
domain between helium and hydrogen burning constitutes the dominant
damping region. Some minor additional damping is observed in the
outermost layers. As the ThMPs begin, the driving in the He-burning
shell overcompensates the intershell damping. Finally, as the star
stabilizes secularly, the driving and damping features persist as
before; however, the damping regions eventually win over the weakening
driving effect in the fattening He shell. For O$_2$ and O$_3$ , only
weak driving occurs in the He-burning shell, the intershell region
exhibits almost canceling driving and damping. It is then the envelope
that eventually determines the fate of the modes' stability with strong
damping throughout most of the outer layers.

\citet{Hansen72} pointed out the running thermal-wave character of 
oscillatory secular modes. Studying the temporal variation of the
complex eigenmodes in our models revealed indeed a pronounced
propagative behavior. All complex eigensolutions behave similarly in
that outward propagating patterns develop in the inter-shell region;
the He-burning shell acts as a~--~spatially stationary~--~piston for
these features. Below the He shell, heat perturbation and radial
displacement behave mostly like standing waves with only a weak
inward-running component.

\subsection{The relevance of the secular modes on the star's evolution}
\label{sect:secevolution}
\begin{figure}
      \resizebox{.94\hsize}{!} {\includegraphics{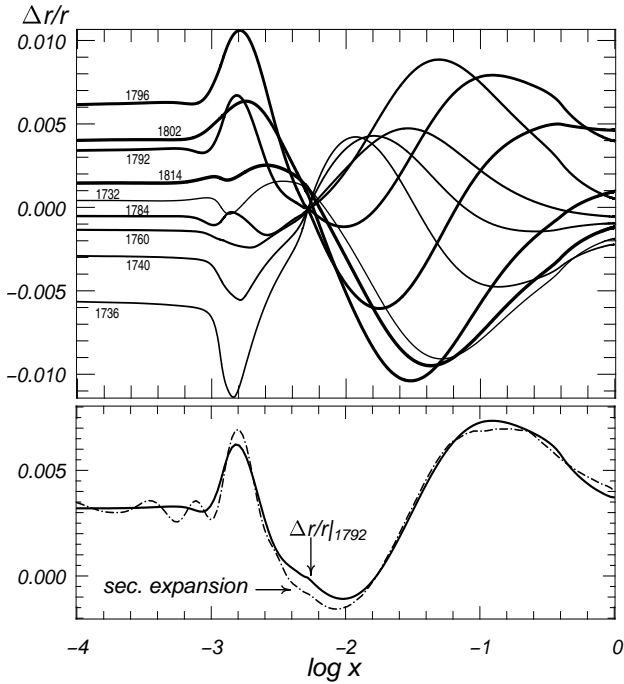}} \caption{
      The top panel shows the relative radius change, $\Delta r / r$,
      as a function of the relative stellar radius $x$ derived from
      the stellar evolutionary sequence, as the star evolves through
      the ThMP-cycle between model numbers $1732$ and $1814$. This
      phase covers the largest-amplitude mode-unfolding episode
      depicted in Fig.~\ref{fig:modiagim30}. The bottom diagram
      compares the quantity $\Delta r / r$ at model $1792$ with the
      result of the Gabriel-expansion with 9 terms, applied to the
      displacement eigenfunction.}
\label{fig:drr_s4}
\end{figure}

At the end, we want to find out how well the
\citet{gabriel72}~--~expansion does in the context of our ThMP problem.
We address this aspect with the help of Fig.~\ref{fig:drr_s4}; its
top panel shows the spatial variation of $\Delta r /r = 1 -
r(m,t_0)/r(m,t_1)$ (with $t_0$, $t_1$ denoting temporally neighboring
models) during the ThMP cycle which is bounded by models $1732$ and
$1814$~--~this cycle encompasses the largest-amplitude mode-unfolding
seen in Fig.~\ref{fig:modiagim30}. The model numbers given in
Fig.~\ref{fig:drr_s4} refer to the later epochs used to compute
$\Delta r /r $; the bump at about $\log x = -2.8$ (location of the
He-shell) is spatially an essentially stationary feature,  the same
applies to the homologous radius variation of the even deeper lying
regions. An almost perfect node (at $\log x \approx -2.2$) lies at the
base of the hydrogen-burning shell. This node is a manifestation of
the ``mirror principle'' which is referred to frequently in the
context of shell burning stars. Finally, an outward propagating
bump develops above the hydrogen-burning shell; its propagation into
the star's envelope can be traced even in the time-series of $\Delta r
/r$ in the top panel of Fig.~\ref{fig:drr_s4}.  

\begin{table*}
\caption{Coefficients of the expansion of $T\diff_t s\vert_{\mathrm
evol}$ in terms of secular eigenmodes for model $1792$. 
The first column gives number
of terms of the series expansion. The modes used in
the series are denoted by O's for the oscillatory ones (cf. 
Fig.~\ref{fig:modiagim30}) and M's for monotonic secular modes. The
second lines in the columns of O$_1$ and O$_2$ give the imaginary
parts of the respective expansion coefficients.  
                                     }
\label{table:ExpandTdS}               
\centering                            
\begin{tabular}{c c c c c c c c c c}  
\hline\hline                          
number of modes & O$_1$ & O$_2$ & O$_3$  
           & M$_1$ & M$_2$ & M$_3$ & M$_4$ & M$_5$ & M$_6$  \\  
\hline                                
3  & 0.177 & 0.301 & -0.031 
   &       &       &        &         &         &       \\
\smallskip
   & 0.427 & 0.124 &   
   &       &       &        &         &         &       
   \\
4  & 0.238 & 0.355 & -0.085 
   &-0.289 &       &        &         &         &       \\
\smallskip
   & 0.513 & 0.270 &   
   &       &       &        &         &         &       
   \\
5  & 0.223  & 0.331 & -0.074 
   &-0.289  & 0.003 &        &         &         &       \\
\smallskip
   & 0.494  & 0.301 &   
   &        &       &        &         &         &       
   \\
6  & 0.264  & 0.363 & -0.087 
   &-0.319  & 0.012 & -0.006 &         &         &       \\
\smallskip
   & 0.536  & 0.236 &   
   &        &       &        &         &         &      
   \\
7  & 0.248  & 0.331 & -0.049 
   &-0.293  & 0.014 & -0.012 & -0.034  &         &       \\
\smallskip
   & 0.456  & 0.360 &   
   &        &       &        &         &         &       
   \\
8  & 0.209  & 0.105 & -0.012 
   &-0.278  & 0.012 & -0.009 & -0.012  & 0.126   &       \\
\smallskip
   & 0.387  & 0.474 &   
   &        &       &        &         &         &       
\\
9  & 0.208  & 0.158 & -0.009 
   &-0.274  & 0.012 & -0.009 & -0.016  & 0.084   & -0.016\\
   & 0.383  & 0.481 &   
   &        &       &        &         &         &       
   \\
\hline                                   
\end{tabular}
\end{table*}

\citet{gabriel72} showed that for sufficiently short time-scales, over
which a star's evolutionary variations can be linearly approximated
relative to a reference epoch, the physical quantities can be expanded
into a series of secular eigenfunctions
\citep[see eq.~(13) in][]{gabriel72}. We followed Gabriel's procedure to
find out how well the linear eigensolutions approximate the variation
induced by the star's short-term evolution. At (between 100 and 300)
equally spaced gridpoints in $\log x$, we solved the least-square
problem for the coefficients of the series expansion in $T\,\delta s$
\citep[cf. eq.~(15) in][]{gabriel72}. Table~\ref{table:ExpandTdS}
lists the results for model $1792$ for various choices of terms
included in the series (first column). For each number of modes
included in the series expansion, the sum of the coefficients was
normalized to unity separately.  The mode designated by O$_3$ was
unfolded at model $1792$ so that its expansion coefficient is real and
not complex as for O$_1$ and O$_2$. The monotonic modes which were
included in the series expansion are labeled as M$_i$. In contrast to
Gabriel's gravitationally contracting model stars, our ThMP models do
not show an equally rapid convergence of the series as a function of
the number of modes (i.e. terms) included.  In accordance with the
eigenmode properties discussed in Sect.~\ref{sect:eigenspectrum} we
found modes O$_1$ and O$_2$ to be essential to approximate well the
functional behavior in the nuclear burning shells; to also match the
core and the envelope variation, the inclusion in the series expansion
of (damped) monotonic modes, in particular of M$_1$, turned out to be
indispensable; however, at the price of added spatial small-amplitude
oscillations in $-4 < \log x < -3$.  An impression of the quality of
the result of Gabriel's procedure applied to $\Delta r /r$ of model
$1792$ with $9$ terms in the series expansion can be got from the
bottom panel of Fig.~\ref{fig:drr_s4}.

The \citet{gabriel72}~--~expansion turned out to do reasonably
well to approximate the spatial variation of perturbed physical
quantities (see bottom panel of Fig.~\ref{fig:drr_s4}).  However, in
contrast to many stellar pulsation problems, linear secular theory
failed to reproduce the amplitude evolution of the variability
throughout the ThMP episode. This is not further surprising as this aim
lies outside the domain of validity of the
Gabriel-expansion. Nevertheless, the period of the overstable secular
mode during the ThMP phase provides a good approximation to the
pulse-cycle length of the fully nonlinear problem. 

Finally, it is worth emphasizing that the secular problem is strongly
non-normal. So that we cannot, in general, expect a close match of the
linear results with the nonlinear developments. For non-normal
operators, considerable transient phases are usually encountered at
the onset of instabilities, so that normal-mode eigenanalyses might
have only limited predictive power
\citep[e.g][]{TrefethenEmbree, Eisenman05}. In our case here, though,
the agreement is surprisingly good.

\section{Conclusions}
\label{sect:conclusio}
A large number of $3 \msol$ stellar models that exhibit thermal
micropulses at the onset of their double-shell nuclear burning phase
were studied by means of linear secular eigenanalyses.  The
\emph{micro}pulses are considered as physically comparable with the
larger-amplitude thermal pulses of low- and intermediate-mass stars
usually encountered along the advanced AGB.  Our computations showed
that the ThMP episode sets in via the instability and finally dies
away via the restabilization of an \emph{oscillatory} secular
eigenmode. And most importantly, the overstability of the
lowest-frequency oscillatory secular mode persists over most of the
ThMP phase. During the numerous ThMP cycles, the oscillatory modes
unfold temporarily into pairs of purely monotonous secular modes. This
phenomenon, we think, is the result of performing stability analyses
on the actually pulsing stellar models.  We conclude that the ThMPs
are the nonlinear development of the linear instability of the
\emph{oscillatory} O$_1$ mode. Most importantly, the 
cyclic nature of the ThMPs has its origin in an unstable linear
oscillatory secular mode that persists throughout the ThMP episode; it
is therefore not surprising that the linear periods
(cf. Fig.~\ref{fig:periods30d} and in particular
Fig.~\ref{fig:supprsspulses}) correlate well with the lengths of the
nonlinear pulse cycles. The view of \citet{defouw73} of the
thermal pulses to be nonlinear relaxation oscillations of monotonic
type appears to have been premature as it relied probably on the
insufficiently long simulation computations of
\citet{haerschild72}. Would the latter authors have had at their
disposal data covering more thermal-pulse cycles, we are confident
that they would have observed a re-merging of the two monotonic mode
branches into a complex pair during later phases of pulse cycles with
sufficiently low amplitude (cf. Fig.~\ref{fig:modiagim30}).  The
linear aspect of the instability which was emphasized in this paper is
not capable to capture the nonlinear, finite-amplitude development of
the various physical quantities during the ThMPs. Indications that
nonlinear processes are indeed important are hinted at in
Fig.~\ref{fig:thmpshells} which shows that the amplitudes of the
radius and density variations in the nuclear-burning shell centers
saturate after about the first four ThMP cycles and in
Fig.~\ref{fig:modiagim30} with the amplitude of $L_{\rm He}$
decreasing slowly after about 397 My despite the persistent
overstability of mode O$_1$ until about 403 My.  

Applying the explanation of thermal pulses as the nonlinear
development of a linearly overstable secular mode leads also to a
natural explanation of the origin of the
core-mass~--~interflash-period relation which was derived along the
advanced AGB \citep{paczynski75}. We conjecture that the
interflash-period is the period of the overstable secular mode which
gets shorter with increasing core mass. Unfortunately, the
evolutionary phase of the ThMPs is so short that the core mass does
not change significantly; therefore, our conjecture could not yet be
underpinned quantitatively. Nevertheless, Fig.~9 hints at a decreasing
period of the overstable secular mode as the star model ascends the
AGB. A detailed study of thermally pulsing AGB stars remains to be
done. 

Well known, but frequently not paid sufficient attention to is the
fact that stellar-evolution results depend, partly even sensitively,
on the time-stepping of their computation. As known for a long time
\citep[e.g][]{rose66}, we could essentially suppress the pulses when 
adopting even a moderately large time-step (a factor less than two
compared with the pulse-resolving time-step). The nonlinear
``pulse-suppressing'' evolution computation showed a very weak
signature in $L_{\rm He}$ of three strongly damped thermal pulses
only. The linear stability analysis on these models continued, on the
other hand, to reveal an unstable oscillatory secular mode with the
instability extending over essentially the same evolutionary phase as
found in the ``non-suppressing'' computation. Hence, the stellar
evolution equations as discretized in most computer codes are
inherently dissipative and they are prone to suppressing secular
instabilities at large time-steps. Therefore, depending on the
particular goal, it is not advisable to eagerly aim at evolving stars
with a minimum of computational time-steps; it also seems dangerous to
estimate the stability of shell sources via local criteria alone in
stellar-evolution computations (cf. Sect.~\ref{sect:localcriteria}).

The possibility to decide observationally if ThMPs are realized in
nature seems bleak; the luminosity variations during the pulses are
very small and on much too long a time-scale to be potentially
detectable \emph{in flagranti}. Furthermore, the number of stars in
the appropriate evolutionary stage in a stellar aggregate such as a
cluster seems too small to be observable via an accumulation effect in
say a color-magnitude diagram. Last but not least, we found no
evidence of interpulse mixing in the envelopes of ThMP-ing stars so
that spectroscopic forensics to identify such a phase appears
implausible.  Nonetheless, from the point of view of
understanding stars, the ThMPs constitute a beneficial computational
laboratory to study fundamental aspects of secular stability of
complex stellar models in advanced evolutionary stages pertinent also
to the nature of thermally pulsing upper-AGB stars.  

\begin{acknowledgements}

     This study was initiated by H. Harzenmoser's inquisitive studying
     and digesting the (also older) stellar-physical literature;
     A.G. is grateful for his persevering stimulus and culinary
     support over all the past years. This research has made use of
     NASA's Astrophysical Data System Abstract Service. The
     authors are most grateful for M. Gabriel's thorough refereeing
     efforts which led to significant improvements and eventually to a
     more coherent presentation of the material.  
\end{acknowledgements}

\appendix

\section{Approximate stability analysis of stars in thermal imbalance}

In the following, the linear equations are compiled as they were used to
compute the radial secular modes of stars in thermal imbalance. We
relied on the perturbed Lagrangian quantities
\begin{equation}
\zeta = \frac{\delta r}{r}, \quad
p = \frac{\delta P}{P}, \quad
t = \frac{\delta T}{T}, \quad
l = \frac{\delta L}{x^2 L_\ast}\,.
\end{equation}
The time dependence of the problem is factorized by $\exp(\imag
\sigma t)$. The eigenvalue of the resulting linear problem, $\sigma$,
is normalized by $\|\sigma\| = \signorm$.  The linearized stellar
structure equations for radiative regions, including approximated
thermal-imbalance terms, are cast into the form
\begin{equation}
x\,\diff_x
   \left( 
   \matrix{
            \zeta \cr
            l     \cr
            p     \cr
            t    
   }\right)  =
\left( \matrix{
         -3  &  0     & -\alpha    & \delta      \cr
          0  & -2     &  s_{23}    &  s_{24}     \cr
      s_{31} &  0     &  V         &   0         \cr
      s_{41} & s_{42} & s_{43}     &  s_{44}     \cr
             }\right) \cdot
   \left(
   \matrix{
            \zeta \cr
            l     \cr
            p     \cr
            t    
         }\right) \,,
\end{equation}
with the secondary quantities in the matrix defined as
\begin{eqnarray}
  s_{23} &=& C_1\left( C_2\,\varepsilon_P + T_P + \imag \sigma
                                  \nabla_{\rm ad} \right) \,,    \\
  s_{24} &=& C_1\left( C_2\,\varepsilon_T + T_T- \imag 
                                          \sigma \right) \,, \\
  s_{31} &=& \left( 4 + 3 c_1 \sigma^2 \right) V \,, \\
  s_{41} &=&  4 V\,\nabla_{\rm rad} \,,\\
  s_{42} &=& -V\,\nabla_{\rm rad}\,x^2 \left(L_\ast/L_r\right) \,, \\
  s_{43} &=& -V\,\nabla_{\rm rad}\,\varkappa_P \,, \\
  s_{44} &=&  V\,\nabla_{\rm rad}\,\left(4 - \varkappa_T\right)   \,,      
\end{eqnarray}
with $c_1 = (r/R_\ast)^3/(M_r/M_\ast)$ and $x = r/R_\ast$. Furthermore,
\begin{eqnarray}
C_1 &=&\frac{M_r U \cp T \|\sigma\|}{L_\ast x^2}   \,,      \\
C_2 &=&\frac{\varepsilon}{\cp T \|\sigma\|}        \,,      \\
C_3 &=&\frac{\diff_t \log T}{\|\sigma\|}           \,,    \\
C_4 &=&\nabla_{\rm ad}\frac{\diff_t \log P}{\|\sigma\|} \,.
\end{eqnarray}
The quantities $U$ and $V$ represent the two homology invariants.
The temporal derivatives of $\log P$ and $\log T$ are the ones used in 
the gravitational-energy term of the stellar-evolution computations. 
Finally, we approximate the thermal-imbalance terms as
\begin{eqnarray}
T_P &=& - C_3 c_{PP} + C_4 \left( \delta_P - \alpha \right) \,, \\
T_T &=& - C_3 c_{PT} + C_4 \left( \delta_T + \delta \right) \,.
\end{eqnarray}
With $c_{Pq}$ we refer to $\partial \log \cp / \partial \log q$, the
same applies to $\delta_q$. The remaining not explicitly defined
variables have either their canonical meaning as used in stellar
structure theory \citep[see e.g.][]{kw} or as referred to in
\citet{gagla90a}. Regularity constraints in the star's center 
enforce the central boundary conditions; at the stellar surface, the
physically motivated ones used in \citet{gagla90a} were implemented.
\bibliographystyle{aa}
\bibliography{StarBase}             

\end{document}